\documentclass[twocolumn,prl,preprintnumbers,superscriptaddress,showpacs]{revtex4}

\usepackage{graphicx}
\usepackage{amsmath,amsfonts,amssymb,amsxtra}
\usepackage{units,formula}

\makeatletter
\DeclareRobustCommand{\chemical}[1]{%
  {\(\m@th
   \edef\resetfontdimens{\noexpand\)%
       \fontdimen16\textfont2=\the\fontdimen16\textfont2
       \fontdimen17\textfont2=\the\fontdimen17\textfont2\relax}%
   \fontdimen16\textfont2=2.7pt \fontdimen17\textfont2=2.7pt
   \mathrm{#1}%
   \resetfontdimens}}
\DeclareRobustCommand{\bchemical}[1]{%
  {\(\m@th
   \edef\resetfontdimens{\noexpand\)%
       \fontdimen16\textfont2=\the\fontdimen16\textfont2
       \fontdimen17\textfont2=\the\fontdimen17\textfont2\relax}%
   \fontdimen16\textfont2=2.7pt \fontdimen17\textfont2=2.7pt
   \mathbf{#1}%
   \resetfontdimens}}
\makeatother

\newcommand{\casrruo}{\chemical{Ca_{2-x}Sr_xRuO_4}}

\newcommand{\srruo}{\chemical{Sr_2RuO_4}}

\newcommand{\lacuo}{\chemical{La_2CuO_4}}
\newcommand{\lasrcuo}{\chemical{(La,Sr)_2CuO_4}}

\newcommand{\lsco}{La$_{2-x}$Sr$_{x}$CuO$_4$}

\newcommand{\nccoef}{Nd$_{1.85}$Ce$_{0.15}$CuO$_4$}

\newcommand{\bkbo}{Ba$_{1-x}$K$_x$BiO$_3$}

\newcommand{\dxz}{\chemical{d_{xz}}}
\newcommand{\dyz}{\chemical{d_{yz}}}
\newcommand{\vq}{\chemical{{\bf q}}}

\newcommand{\vQ}{\chemical{{\bf Q}}}

\newcommand{\ozz}{\chemical{{O_{zz}}}}

\newcommand{\kommentar}[1]{}

\begin{document}

\title{Lattice dynamics and electron-phonon coupling in  Sr$_2$RuO$_4$}

\author{M. Braden}
\email{braden@ph2.uni-koeln.de}%
\affiliation{II. Physikalisches Institut, Universit\"at zu K\"oln,
Z\"ulpicher Str. 77, D-50937 K\"oln, Germany}

\author{W. Reichardt}
\affiliation{Forschungszentrum Karlsruhe, Institut f\"ur
Festk\"orperphysik, P.O.B. 3640, D-76021 Karlsruhe, Germany}

\author{Y. Sidis}
\affiliation{ Laboratoire L\'eon Brillouin, C.E.A./C.N.R.S.,
F-91191 Gif-sur-Yvette Cedex, France}

\author{Z. Mao}
\affiliation{ Department of Physics, Tulane University, New
Orleans, Louisiana 70118, USA}

\author{Y. Maeno}
\affiliation{Department of Physics, Kyoto University, Kyoto
606-8502, Japan}

\date{\today, \textbf{preprint}}

\pacs{PACS numbers: 74.25.Kc  74.70.Pq}

\begin{abstract}

The lattice dynamics in Sr$_2$RuO$_4$ has been studied by
inelastic neutron scattering combined with shell-model
calculations. The in-plane bond-stretching modes in Sr$_2$RuO$_4$
exhibit a normal dispersion in contrast to all electronically
doped perovskites studied so far. Evidence for strong electron
phonon coupling is found for c-polarized phonons suggesting a
close connection with the anomalous c-axis charge transport in
Sr$_2$RuO$_4$.

\end{abstract}

\maketitle

\section{Introduction}

The discovery of superconductivity in \srruo \ \cite{1} has
attracted attention partially due to the structural similarity
with the high-temperature superconducting cuprates.
Superconductivity in \srruo \ is unconventional in character: from
Knight-shift, $\mu$SR and polarized neutron studies a triplet
pairing was deduced \cite{maeno-mackenzie}. Recently, direct
evidence for an odd pairing symmetry was obtained by experiments
on quantum interference devices \cite{liu} and by magneto-otpical
Kerr effect \cite{kerr}. In view of the unconventional pairing
symmetry, the conventional electron-phonon coupling can be
discarded from the possible pairing mechanisms, however, one may
not rule out some exotic electron-lattice mechanism to play an
important role. Indeed, the superconducting transition in \srruo
\ exhibits a remarkable isotope effect \cite{mao}. A detailed
study of the lattice dynamics in this material appears hence
highly desirable.

The lattice dynamics in \srruo \ merits further interest as a
reference material for the large class of transition metal-oxides
with a perovskite related structure where an anomalous phonon
dispersion has been observed
\cite{4,5,6,6a,7,8,8a,8b,26,27,15,21,14,12,19,20}. Stimulated by
the phenomenon of high-temperature superconductivity the phonon
dispersion of many oxide perovskites has been studied in detail.
In general the lattice dynamics of these materials is well
understood with the aid of ionic lattice-dynamical models
\cite{chaplot,bruesch}. In particular in the insulating parent
compounds, like \lacuo , all features in the phonon dispersion
are at least qualitatively described by simple ionic models. In
electronically doped materials, like \lasrcuo , however,
significant discrepancies in the dispersion of the longitudinal
bond-stretching branches were discovered
\cite{4,5,6,6a,7,8,8a,8b,26,27,15,21,14,12,19,20}. A typical
lattice dynamical model including an isotropic screening function
will always predict an increasing bond-stretching dispersion when
passing from the Brillouin-zone center into the zone due to the
imperfect screening of the Coulomb-potentials at short distance.
Instead a decreasing dispersion with the longitudinal
zone-boundary frequencies falling below those of the transverse
branch has been observed in numerous perovskite materials : in
different types of superconducting cuprates
\cite{4,5,6,6a,7,8,8a,8b,26,27}, in nickelates \cite{15,21}, in
manganates \cite{14} and in superconducting \bkbo
\cite{12,19,20}, for a recent review of these effects see
reference \cite{27}. These doping-induced effects are frequently
called over-screening \cite{tachiki} referring to the fact that
simple screening should soften the longitudinal phonon modes. All
these materials, however, are poor metals and they are all close
to a charge ordering instability \cite{imada}. One may, hence,
consider the frequency softening in the bond-stretching
dispersion as a dynamic precursor of the charge ordering. Charge
ordering at the metal site will modulate the bond-distances to
the O-surroundings \cite{brown}. Therefore, charge ordering and
the bond-stretching phonons are intimately coupled. For instance
in the cuprates, the propagation vector where static stripe
ordering is observed upon Nd-codoping agrees well with the \vq
-vector of the anomaly in the longitudinal bond stretching branch
\cite{6}. A pronounced anomaly has recently been observed in
materials passing into the charge-ordered stripe phase
\cite{reznik}.

As all the electronically doped oxide perovskites studied so far
exhibit an anomalous bond-stretching dispersion, one may ask
whether the over-screening effect should be considered as being
anomalous. The study of additional materials is hence needed.

We have studied the phonon dispersion in \srruo \ by inelastic
neutron scattering and we have modeled it with ionic shell models
taking the screening of the Coulomb potentials into account. The
over-screening effect in the in-plane longitudinal
bond-stretching dispersion is not found in \srruo \ which is the
only metallic oxide perovskite exhibiting a normal
bond-stretching dispersion. However, the inter-plane charge
dynamics seems to induce a pronounced softening of the $c$-axis
polarized bond-stretching modes which must be taken as evidence
for a strong-electron phonon coupling in this unconventional
superconductor.

\begin{figure}
\includegraphics[width=0.46\textwidth]{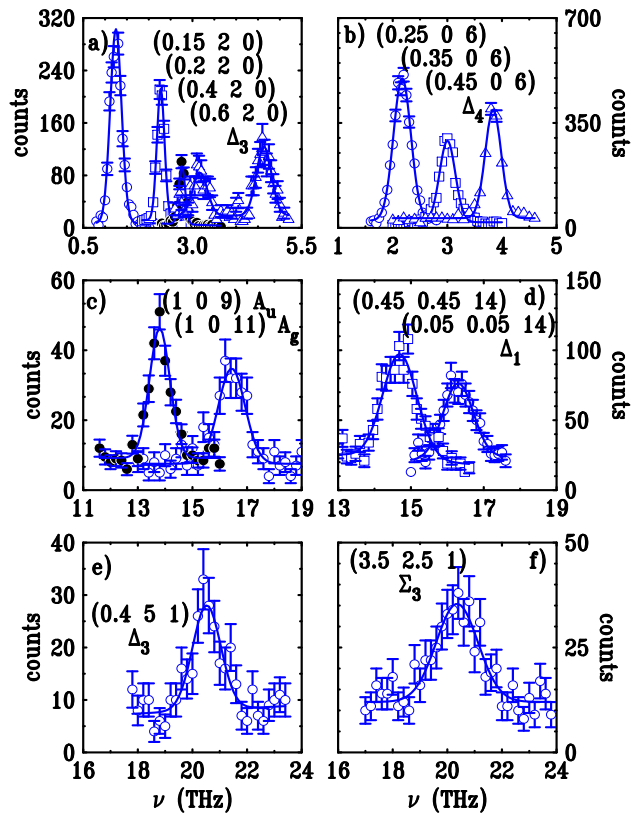}
\caption{(color online) Typical scans to determine the phonon
dispersion in \srruo . In parts a) and b) transverse acoustic
phonon modess of $\Delta _3$ and $\Delta _4$ symmetry are shown,
respectively. In part c) scans across the two $c$-polarized
apical-oxygen modes, $A_u$ and $A_g$, are presented. At the $A_g$
mode a $\Delta_1$ branch starts for which part d) shows two
corresponding scans. In parts e) and f) scans across transverse
in-plane bond-stretching modes are shown : e) for a transverse
mode along the [100] direction and f) for the zone-boundary mode
in [110] direction which is the quadrupolar mode. } \label{scans}
\end{figure}

\section{Experimental}

The single crystals studied in this study were grown by the
floating-zone technique \cite{crystal-growth} and these and
crystals grown in the same way were already used in numerous
inelastic neutron-scattering experiments aiming at the magnetic
excitations in \srruo \ \cite{sidis,braden2002} as well as in many
other studies \cite{maeno-mackenzie}. Measurements were performed
on the 1T thermal triple-axis spectrometer at the Orph\'ee reactor
in Saclay. In the standard configuration we have used a
pyrolithic graphite (PG) (002)-Bragg reflection monochromator at
low energies (up to $\sim$10\ THz energy transfer) and a Cu-(111)
monochromator at high energy. In order to suppress higher-order
contaminations we have set a PG filter into the scattered beam and
fixed the final energy to 3.55 or 7.37\ THz (corresponding to
14.7 and  30.5\ meV). The crystals were cooled to about 15\ K with
the aid of a closed-cycle cryostat. Most of the measurements were
taken at that temperature. In order to study specific temperature
dependencies a helium cryofurnace was used.

Typical scans aiming at phonons at quite different energies are
shown in Fig. 1 covering the full energy range of the phonon
dispersion in \srruo . In an inelastic neutron scattering
experiment the scattering intensity, $I$, arising from a
one-phonon process in neutron-energy loss mode is given by
\cite{squires}:

$$
I \propto  {1\over \omega} \cdot (n(\omega)+ 1) \{\sum_{d}{b_d
\over \sqrt{m_d}} \cdot e^{(-W_d+i{\bf Q}\cdot {\bf r_d})}
\cdot({\bf Q}\cdot {\bf e_d})\}^2  ~~~~~~(1),$$

($\omega$ denotes the phonon frequency ($\omega =2\pi \nu$, note
that 1\ THz corresponds to 4.14\ meV), ${\bf q}$ the phonon
wave-vector, ${\bf Q}={\bf g} +{\bf q}$ the scattering vector
(${\bf g}$ the reciprocal lattice vector) $e^{-W_d}$ the
Debye-Waller-factor, and the sum extends over the atoms in the
primitive cell with mass $m_d$ scattering length $b_d$, position
${\bf r_d}$ and polarization vector ${\bf e_d}$. In general,  the
intensity is determined by the Bose-factor, $n(\omega)$, and the
$1\over \omega$-term, which strongly reduce the effectiveness to
observe high-energy modes. The dynamic structure factor equation
(1) allows one to determine the full polarization of the modes by
comparing the intensities measured in different Brillouin zones.
In the other sense, a preliminary model may predict via equation
(1) the favorable Brillouin-zones for the observation of
particular phonon modes. Throughout this work we were guided by
the lattice-dynamical model, which was successively improved with
the growing amount of experimental information.

\section{Overall phonon dispersion and lattice dynamical model}

\begin{figure}
\widetext
  \includegraphics*[width=0.93\textwidth]{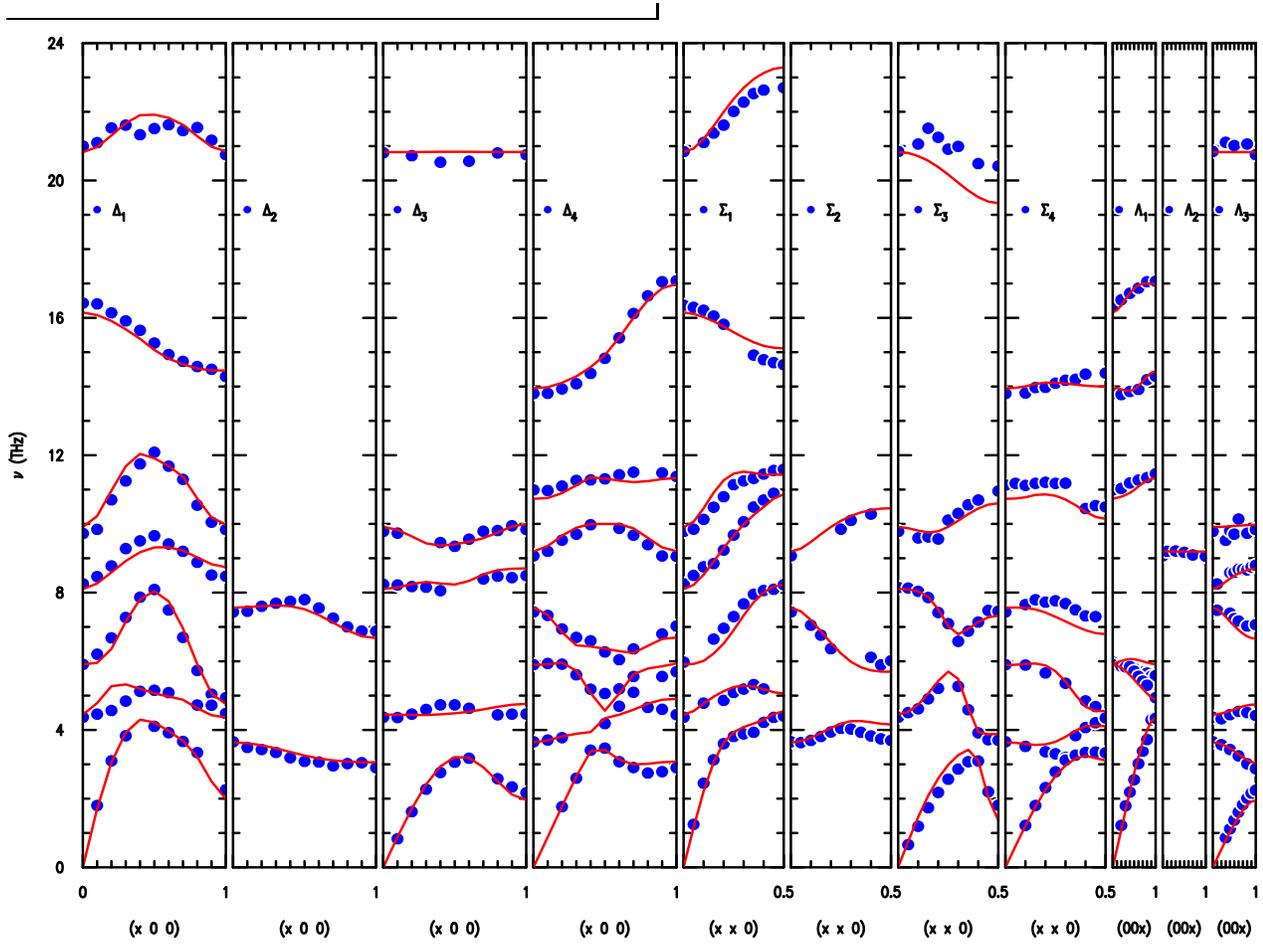}
 \caption{(color online) Phonon dispersion in \srruo , symbols denote the measured
 frequencies and lines those calculated with the lattice dynamical model. We show the phonon
 dispersion along the main symmetry directions, [100], [110] and [001], separated according to the
 irreducible representations, see text and Table I.}\label{dispersion}
\narrowtext
\end{figure}

Fig. \ref{dispersion} shows the measured phonon dispersion
together with the results of the lattice dynamical calculations.
\srruo \ crystallizes in the K$_2$NiF$_4$-structure in space group
$I4/mmm$ with seven atoms in the primitive cell, accordingly the
phonon dispersion consists of 21 phonon modes at the general
position, but due to symmetry there are several degenerations at
special points or lines. A small part of this data was already
published earlier focusing on the lattice instability appearing
at the zone-boundary {\bf M}=(0.5,0.5,0) of the $\Sigma _3$-branch
\cite{phon-sro1}. The mode associated with the rotation of the
RuO$_6$-octahedra around the $c$ axis possesses a rather soft
frequency as seen in the dip in the $\Sigma _3$ dispersion at the
M-point, (0.5,0.5,0),  and this mode exhibits an anomalous
temperature dependency as reported in our first publication
\cite{phon-sro1}. These effects must be considered as a precursor
of the structural phase transition lowering the space-group
symmetry from $I4/mmm$ to $I4_1/acd$ which has been observed in
the meanwhile in the \casrruo -series
\cite{csro-struc1,csro-struc2}. In contrast to the rotational
mode, the octahedron-tilting mode does not show anomalous effects
in \srruo . This dynamic precursor of the rotational instability
is perfectly described by the lattice-dynamics model.

In a first approach, one may attribute the branches in the range
of 4\ THz to Sr/Ru-vibrations and those in the range 5-12\ THz to
different Ru and Ru-O bond-bending modes. The modes in the
frequency range 14-16\ Thz correspond to vibrations of the apical
oxygen parallel to the $c$ axis and the highest band of phonons
with frequencies above 20\ THz is associated with the in-plane
bond-stretching vibrations. Only three zone-center modes are
Raman active and were observed by Udagawa et al.\
\cite{udagawa1,udagawa2}. The two $A_g$-modes corresponding to
the $c$-axis vibrations of Sr and to that of the apical oxygen are
found at 5.96 and 16.49 \ THz, respectively, and one $E_g$ mode
is found at 7.36\ THz \cite{udagawa1,udagawa2}. The corresponding
frequencies determined by inelastic neutron scattering perfectly
agree with these Raman results, we find the frequencies 6.02,
16.44 and 7.27\ THz. Due to the good in-plane electrical
conductivity, Infra-Red optical measurements with the polarization
parallel to the planes were unable to observe any phonon modes but
the three c-polarized $A_u$ frequencies at 5.92, 11.11 and 14.49 \
THz have been determined \cite{katsufuji}. The two lower
frequencies agree well with our results giving 5.94 and 10.99\
THz. However, we find the highest $A_u$ mode at 13.80\ THz. Since
the determination of the phonon frequencies has not been the
focus of the Infra-Red study, no raw-data were shown in ref.
\cite{katsufuji} and we cannot further comment the high-frequency
discrepancy.

The 21 modes at the zone-center can be separated according to the
irreducible representations into : 1\ $B_u$ (observed at 9.09\
THz), 2\ $A_g$ (6.02 and 16.44\ THz), 4\ $A_u$ (5.94, 10.99 and
13.80\ THz plus acoustic mode), 2\ double-degenerated $E_g$ ( 3.59
and 7.27\ THz), and 5\ double-degenerated $E_u$ ( 4.38, 8.36,
9.72, and 20.85\ THz and acoustic mode). Along the main-symmetry
directions [100] ($\Delta$), [110] ($\Sigma$), and [001]
($\Lambda$) the modes can be separated according to three or four
irreducible representations. In our notations the subscript 1
always refers to the representation the longitudinal acoustic mode
belongs to. Subscript 2 signifies the representation which does
not contain any acoustic mode, and subscripts 3 and 4 (the latter
is not used for $\Lambda$ modes) denote the representations
containing the transverse acoustic modes with c-polarized acoustic
modes belonging to representation 4. The polarization schemes and
the compatibility relations of the irreducible representations
are given in Table I.

\bigskip

\begin{table}
\begin{tabular}{| c | c |c |c |c |c  |c  |c  |}
\hline
~~ & Sr & Sr' & Ru  & O1 & O1' & O2 & O2'  \\
~~ & 0\ 0\ .353 & 0\ 0\ -.353 & 0\ 0\ 0 & .5\ 0\ 0 & 0\ .5\ 0&0\ 0\ .162 & 0\ 0\ -.162\\
\hline
1\ B$_{u}$ & 0 0 0 & 0 0 0 & 0 0 0 & 0 0 A & 0 0 -A & 0 0 0& 0 0 0 \\
2\ A$_{g}$ & 0 0 A & 0 0 -A & 0 0 0 & 0 0 0 & 0 0 0 & 0 0 B& 0 0 -B\\
4\ A$_{u}$ & 0 0 A & 0 0 A & 0 0 B & 0 0 C & 0 0 C & 0 0 D& 0 0 D\\
2\ E$_{g}$ & A 0 0 & -A 0 0 & 0 0 0 & 0 0 0 & 0 0 0 & B 0 0& -B 0 0\\
5\ E$_{u}$ & A 0 0 & A 0 0 & B 0 0 & C 0 0 & C 0 0 & D 0 0& D 0 0\\
\hline
7 $\Delta _1$  & A 0 B & A 0 -B & C 0 0 & D 0 0 & E 0 0 & F 0 G& F 0 -G  \\
2 $\Delta _2$   & 0 A 0 & 0 -A 0 & 0 0 0 & 0 0 0 & 0 0 0 & 0 B 0& 0 -B 0 \\
5 $\Delta _3$   & 0 A 0 & 0 A 0 & 0 B 0 & 0 C 0 & 0 D 0 & 0 E 0& 0 E 0 \\
7 $\Delta _4$   & A 0 B & -A 0 B & 0 0 C & 0 0 D & 0 0 E & F 0 G& -F 0 G \\
\hline
7 $\Sigma _1$  & A A B  & A A -B & C C  0 & D E 0 & E D 0 & F F G& F F -G \\
3 $\Sigma _2$  & A -A 0 & -A A 0 & 0 0  0 & 0 0 B & 0 0 -B & C -C 0& -C C 0  \\
5 $\Sigma _3$  & A -A 0 & A -A 0 & B -B 0 & C D 0 & -D -C 0 & E -E 0& E -E 0 \\
6 $\Sigma _4$  & A A B  & -A -A B & 0 0 C & 0 0 D & 0 0 D & E -E F & -E E F \\
\hline
6 $\Lambda _1$  & 0 0 A & 0 0 B & 0 0 C & 0 0 D & 0 0 D & 0 0 E& 0 0 F \\
1 $\Lambda _2$  & 0 0 0 & 0 0  0 & 0 0  0 & 0 0 A & 0 0 -A & 0 0 0& 0 0 0  \\
7 $\Lambda _3$  & A 0 0 & B 0 0 & C 0 0 & D 0 0 & E  0 0 & F 0 0 & G 0 0 \\
\hline
\end{tabular}

\begin{tabular}{ c c c  }
 \hline
 {[100]}   & $\Delta _1$ : 2\ $A_g$ + 5\ $E_u$ & $\Delta_2$ : 2\ $E_g$ \\
  & $\Delta_3$ : 5\ $E_u$ & $\Delta _4$ : 1\ $B_u$ + 2\ $E_g$ + 4\ $A_u$ \\
 {[110]}  & $\Sigma _1$ : 2\ $A_g$ + 5\ $E_u$ & $\Sigma_2$ : 1\ $B_u$ + 2\ $E_g$ \\
 & $\Sigma_3$ : 5\ $E_u$ & $\Sigma _4$ : 2\ $E_g$ + 4\ $A_u$ \\
 {[001]}   &$\Lambda _1$ : 2\ $A_g$ + 4\ $A_u$ &$\Lambda_2$ : 1\ $B_u$           \\
  & 2$\cdot \Lambda_3$ : 2\ $E_g$ + 5\ $E_u$   & / \\
\end{tabular}
\nobreak \caption{Upper part :Polarization schemes according to
the crystal structure of \srruo ~ for all $\Gamma$-modes and for
the representations along the [100], [110] and [001] directions
(labeled $\Delta$, $\Sigma$ and $\Lambda$ respectively); the first
line gives the positions of 7 atoms forming a primitive unit
\cite{braden-struc}, the following lines show the displacements of
these atoms. A letter at the $i$-position, signifies that this
atom is moving along the $i$-direction, a second appearance of
the same letter signifies that the second atom moves with the same
amplitude ("-" denotes a phase shift) in the corresponding
direction. Lower part : Compatibility relations of the
irreducible representations for phonon wave vectors along the
main-symmetry directions. }\label{comp}
\end{table}

\clearpage

\narrowtext

\clearpage

We have tried to describe the observed phonon dispersion of \srruo
\ in a Born-von-Karman force-constant model taking into account
atomic pairs up to 15th shell or 30 parameters. For a metallic
material one may expect such a model to yield a good description.
In \srruo \ there are clear discrepancies, for example the
relatively large elastic constants compared to the lower
zone-boundary frequencies, which suggest the relevance of
long-range interactions arising from Coulomb potentials. We,
therefore, have modeled the phonon dispersion of \srruo \ in an
ionic shell model, whose parameters are given in table I.
Starting from the generalized model presented in reference
\cite{chaplot} we have adapted the parameters to the measured
dispersion under the additional constraint of the resulting
forces not getting too large.

For the description of the phonon dispersion in \srruo \ we use
Coulomb-potentials with effective charges, $V(r)\propto
{Z_1Z_2e^2\over r}$ and the repulsive forces are described by
Born-Mayer potentials $ V(r)=A\cdot exp(-r/r_o)$. Shell charges
describing a single-ion polarizability have been introduced for
all atoms with  shell-core force constants  chosen anisotropic for
Sr and Ru. The interatomic forces act on  the shells in our model.
We treat the in-plane oxygen (O1) and the apical oxygen (O2)
differently, similar to the case of the cuprates, but the same O-O
potential was used for all pairs. For the O-O-interaction we have
added a van der Waals term of -100(ev \AA$^6$)$\cdot r^{-6}$ with
$r$ the interatomic distance in \AA. All calculations were
performed with the GENAX program \cite{genax}.

\bigskip
\begin{table}
\begin{tabular}{c c c c }
\multicolumn{4}{c}{ionic part} \\
\hline
ion   & Z     & Y     & K    \\
Ru    & 2.58  & 0.47   & 8.0/2.0  \\
Sr    & 2.00  &  5.86  & 3.6  \\
O1    & -1.52 & -3.25  & 1.8  \\
O2    & -1.77 & -2.77  &  1.8 \\
\end{tabular}
\begin{tabular}{c c c c}
\multicolumn{4}{c}{potentials} \\
\hline
pair  & A(eV)  & r$_0$ (\AA ) & C (eV/\AA $^6$ \\
Sr-O1  &  1825  & 0.318 & -                  \\
Sr-O2  &  2250  & 0.318 & -                  \\
Ru-O1  &  2999  & 0.260 & -               \\
Ru-O2  &  3874  & 0.260 & -               \\
O1-O1   &  2000  & 0.284 &  -100              \\
O1-O2   &  2000  & 0.284 &  -100              \\
O2-O2   &  2000  & 0.284 &  -100              \\
~ & ~ & ~ & ~ \\
\end{tabular}
\begin{tabular}{c c }
\multicolumn{2}{c}{additional force constants (dyn/cm)} \\
\hline
constant  &  F       \\
angular O1-Ru-O1 & 6798    \\
angular O2-Ru-O2 & 11178    \\
special O2-z & 3205      \\
special breathing & -5203 \\
\end{tabular}
\nobreak \caption{Lattice dynamics model  parameters for the
description of the phonon dispersion in \srruo ; for the
explanation of the parameters see text; Z,Y are given in electron
charges; K in $10^6$dyn/cm. }\label{tab1}
\end{table}

In addition to the potentials a few additional force constants
were introduced to improve the description. Angular forces were
added for the linear O1-Ru-O1 and O2-Ru-O2 bonds and special force
constants were needed in order to describe the behavior of the
c-polarized bond-stretching modes, as it will be discussed below.

\section{Soft rotational mode}

Since our first report on the softening of the rotational mode in
\srruo \ the structural studies on the phase diagram of \casrruo
\ have shown the close connection between the static octahedron
rotation and the electronic and magnetic properties in \casrruo \
\cite{csro-struc1,csro-struc2}. Therefore, it appeared
interesting to study the temperature dependence of the rotational
mode in more detail in \srruo \ which, however, does not exhibit
the static distortion \cite{braden-struc}. The new results on the
rotational mode confirm our previous measurement and are given in
Fig. 3. The rotational mode was measured at \vQ =(2.5,1.5,0) in
the [100],[010] scattering geometry. Between the lowest and the
highest temperature ($\sim$450\ K) the rotational mode
continuously broadens whereas the frequency first softens up to
about room temperature and then stiffens upon further temperature
increase. These temperature dependencies cannot be explained by a
normal structural instability where the mode should soften and
broaden concomitantly. Instead, we think that these anomalous
effects have to be attributed to electron-phonon coupling. The
strongest temperature effects in both, in the frequency and in the
width, are observed near 150\ K, where the $c$-axis resistivity
exhibits a broad maximum. The anomalous temperature dependency of
the rotational mode might thus be connected with the $c$-axis
charge transport.

\begin{figure}
  \includegraphics[width=0.45\textwidth]{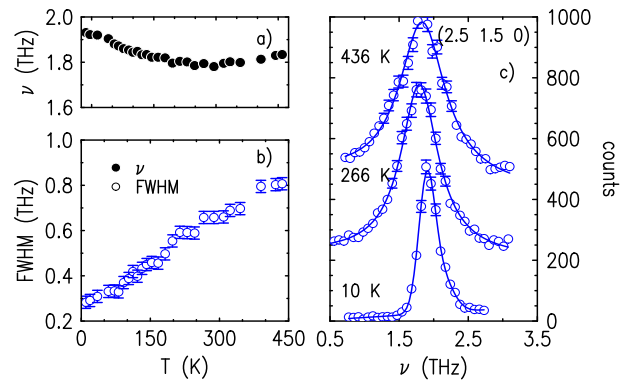}
  \caption{(color online) Temperature dependence of the rotational
  mode in \srruo , a) mode frequency and b) width. The mode was measured
  at \vQ =(2.5,1.5,0) in the [100],[010] scattering geometry. Raw-data scans
  at three temperature are shown in part c), these scans were fitted by Lorentzian
  profiles in order to obtain the positions and widths. }
  \label{rotmode}
\end{figure}

\section{Screening of Coulomb potentials by charge carriers}

The metallic character of \srruo \ yields an effective screening
of the Coulomb potentials and was  explicitly taken into account
in the lattice-dynamics calculations. We describe the electronic
susceptibility by the Lindhard function for a free-electron gas
\cite{chaplot}. In such model the screening is perfect at long
distance, thereby the splitting between longitudinal optic (LO)
and transverse optic (TO) phonon frequencies disappears at the
Brillouin-zone center. However, when passing into the zone, the
screening will be less effective as it has to occur at shorter
distance and the LO-TO splitting may partially recover yielding
an increasing dispersion close to the zone center. The screening
scenario applies to any polar modes, in particular to those with
Ru-O bond-bending character, which posses the strongest polar
character in \srruo . Since in other perovskites these bending
modes are associated with ferroelectricity, they are frequently
called ferroelectric modes. Comparison of the \srruo \
low-temperature phonon dispersion with those observed in
La$_{1.85}$Sr$_{0.15}$CuO$_4$ \cite{chaplot}  and in \nccoef \
\cite{27} immediately reveals differences. The bond-bending
branches in the ruthenate are much flatter indicating a more
effective screening. The relevant $\Delta _1$ and $\Sigma _1$
branches starting at the $E_u$ mode at 9.72 \ THz attain only
frequencies of about 12\ THz in the ruthenate whereas they
exhibit a steep dispersion up to 16\ THz in the cuprates starting
at a comparable zone-center frequency. Also along the $c$
direction the screening is rather perfect in \srruo , as the
$\Lambda _1$ branch starting at the 11\ THz $A_u$ mode is rather
flat. The comparably efficient screening perpendicular to the
planes is further corroborated by the model calculations where an
isotropic screening formalism is found to describe the dispersion
relation reasonably well. The isotropic screening
 contrasts with the anisotropic transport properties of
\srruo , where resistivity ratios ${\rho _c \over
\rho_{in-plane}}$ of the order of 1000 were reported
\cite{hussey,tyler}.

In the case of \srruo \ we obtain a screening vector of $k_s$ =
0.86\AA$^{-1}$ which is considerably larger than the one  found
for example in \lsco , $k_s$ = 0.39\AA$^{-1}$ for x$\sim$0.1
\cite{chaplot} or the one observed in Nd$_{2-x}$Ce$_{x}$CuO$_{4}$
\ $k_s$=0.41\AA$^{-1}$ for x=0.15 \cite{27}. Qualitatively, the
efficient screening agrees with the better metallic properties of
the ruthenate. Using a screening vector of $k_s$=0.42\AA$^{-1}$,
similar to those observed in the cuprates, we find also a
comparable phonon dispersion. The longitudinal branches starting
at the ferroelectric bond-bending modes are then found to attain
frequencies of 16\ THz for propagation vectors along and
perpendicular to the planes as it is indeed observed in the
cuprates.

At low temperature, \srruo \ exhibits a considerably higher
in-plane electrical conductivity than the cuprates. However, upon
heating the conductivity considerably decreases in \srruo \
attaining values at high temperature which are comparable to
those observed in the cuprates. The in-plane mean free path in
\srruo \ was reported to fall below 1 \AA \ without any sign for
resistivity saturation upon heating till 1200\ K \cite{tyler}.
One may, therefore, wonder, whether the efficient low-temperature
screening in \srruo \ is significantly reduced upon heating as
well. To study this problem we have analyzed the ferroelectric
bond-bending modes at different temperatures for the $\Delta _1$
and $\Lambda _1$ branches. However, all temperature-induced
phonon-frequency shifts are very small, of the order of one
percent, whereas an essential change of the screening rendering
the ruthenate at room temperature comparable to the cuprates
would imply frequency shifts by more than 30 percent. We may
conclude that the high screening of the Coulomb potentials in
\srruo \ remains essentially unchanged between 10\ K and room
temperature. The screening of Coulomb potentials seems to be
fully decoupled from the macroscopic charge transport.

Qualitatively the minor temperature-induced frequency-changes
agree with a very small reduction of the screening upon heating;
the zone-center modes slightly soften ($E_u$) or remain unchanged
($A_u$) whereas there is a minor hardening for polar modes in the
Brillouin zone upon heating, but none of these effects is stronger
than two percent.

\section{Dispersion of bond-stretching modes}

The dispersion of the in-plane bond-stretching modes is shown in
detail in Fig. 4. Starting at the highest $E_u$ frequency a
$\Delta_1$ branch connects to the half-breathing mode at \vq
=(0.5,0,0) and then continues to the ${\bf Z}$=(1,0,0) point
bond-stretching mode with an out-of-phase vibration of
neighboring planes. For the corresponding polarization patterns
see Fig. 5a) and b). Unfortunately, there is a strong loss in the
dynamical structure factor when approaching the half-breathing
mode rendering these measurements always difficult. The raw data
shown in Fig. 4a), however, document that the branch can be
followed along this path by combining measurements in different
Brillouin zones, mostly (4,0,0), (5,1,0) and (5,0,3). This branch
shows a rather flat dispersion and a nearly \vq -independent
width. There is no overscreening effect along the [100]-direction
comparable to those observed in all metallic perovskites observed
so far. The dispersion is reasonably well described by our
lattice-dynamical model, see Fig. 4c).

Also along the [110]-direction, we do not find the over-screening
effect but a rather normal and increasing dispersion towards the
planar breathing mode at \vq =(0.5,0.5,0), for the polarization
pattern see Fig. 5c). Fig. 4b) shows the scans taken most at \vQ
=(3+$\xi$,3+$\xi$,2) and in Fig. 4d) the dispersion is presented,
which again is well reproduced by the shell model.

The normal bond-stretching dispersion contrasts with all
electronically doped perovskite compounds studied so far, as all
of them exhibit a pronounced over-screening effect at least along
one of the in-plane directions. The difference between \srruo \
and the materials studied previously consists in the fact that
\srruo \ is intrinsically metallic implying a relatively higher
charge carrier density. More importantly, it appears that the
charge carriers in \srruo \ possess a higher mobility. \srruo \
exhibits a low residual resistivity and well-defined Fermi-liquid
behavior at low temperatures \cite{maeno-mackenzie}. The
ruthenate does not seem to be close to any charge-ordering
instability with an inhomogeneous distribution of charges within
the RuO$_2$-layers.

The debate about the role of electron-phonon coupling in cuprate
high-temperature superconductivity has gained further weight with
the interpretation that the kink observed in ARPES spectra of
many HTSC compounds arises from electron-phonon coupling
\cite{lanzarra}. This interpretation contrasts with the
explanation given by other groups, who attribute the kink to an
interaction with magnetic excitations \cite{dordevic}. A
comparable kink has more recently been found in the ARPES spectra
of \srruo \ as well \cite{aiura}. Therefore, the comparison of the
lattice dynamics of this material with that of the cuprates might
help understanding the origin of these kinks. In view of the
absence of the corresponding phonon anomaly in the in-plane
bond-stretching dispersion of the ruthenate it appears rather
unlikely that the in-plane bond-stretching modes are essential
for the understanding of the ARPES kinks. Instead the ARPES kinks
could arise from the magnetic excitations in both materials as
\srruo \ and \lasrcuo \ exhibit both magnetic fluctuations of
nearly antiferromagnetic character and of comparable strength
\cite{sidis,braden2002,servant}.

\begin{figure}
  \includegraphics[width=0.45\textwidth]{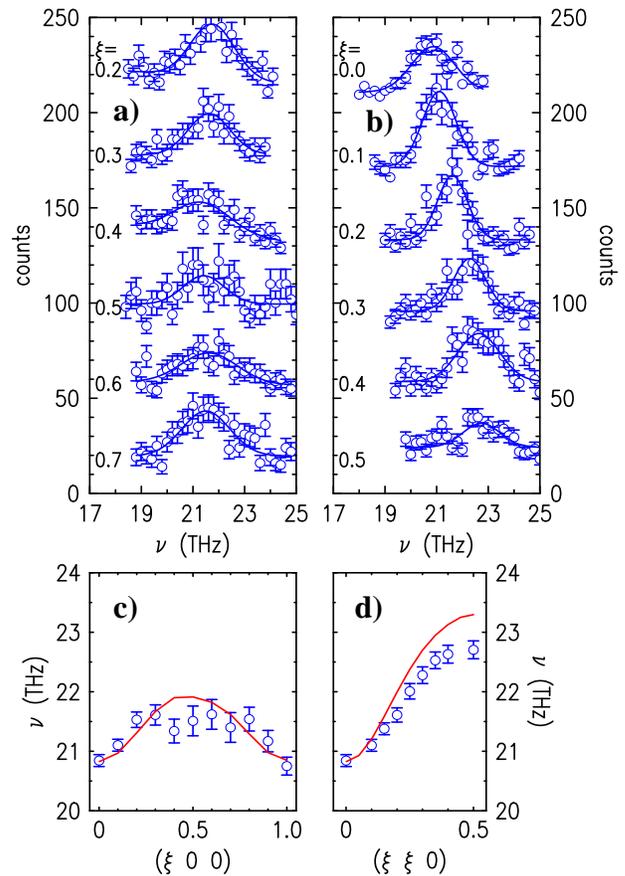}
  \caption{(color online) Raw data scans to determine the
  longitudinal bond-stretching modes at \vq =($\xi$,0,0) a) and at
  ($\xi$,$\xi$,0) b), for clarity scans were successively
  translated vertically.
  Dispersion of the in-plane bond-stretching phonon modes along the
  [100] c) and [110]-direction d);
  symbols denote the measured frequencies
  and lines the dispersion calculated
  with the lattice dynamics model. }
  \label{bonstr}
\end{figure}

Fig. 5 e-g) show the polarization patterns of different
out-of-plane bond-stretching modes. At the zone-center there is
an odd $A_u$ and an even $A_g$-mode where the two O2's connected
to a Ru-site vibrate in and out-of-phase, respectively. Due to
the translation symmetry, the octahedron at the center of the
tetragonal cell moves in phase with the one at the origin for the
zone-center modes. For \vq \ at the point {\bf Z}=(1,0,0),
however, the two octahedra vibrate with opposite phases, see Fig.
5. The $A_g$-mode connects with the so-called \ozz -mode at {\bf
Z} through a $\Delta _1$ branch and the $A_u$-mode to its
analogue through a $\Delta _4$-branch. The \ozz -mode is
interesting in terms of a possible coupling with charge
fluctuations. At the same time all oxygens connected to a certain
Ru-layer move towards that layer whereas those O2's connected to
the neighboring layers move away from those. This mode may,
hence, be considered as the $c$-axis polarized analogue of the
linear breathing mode; it can be coupled with an inter-layer
charge transfer.

\begin{figure}
  \includegraphics[width=0.45\textwidth]{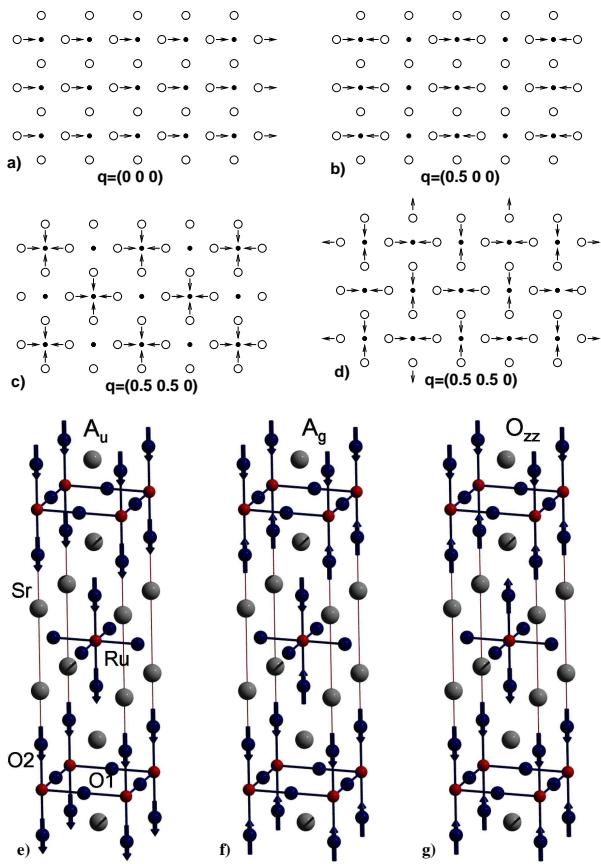}
  \caption{(color online) Polarization patterns of several bond-stretching modes. Parts a), b)
  and c) display the elongations within a single RuO$_2$ layer for the in-plane bond-stretching
  modes at the zone center, at \vq=(0.5,0,0) (half-breathing mode) and at
  \vq =(0.5,0.5,0) (planar beathing mode), respectively.
Part d) shows the polarization pattern of the transverse
bond-stretching mode at \vq=(0.5,0.5,0) of quadrupolar character.
Three z-polarized bond-stretching mode patterns are given in
parts e)-g) :
  the $A_u$ (e) and the $A_g$ (f) modes at the Brillouin-zone center and the \ozz \ mode at the
  {\bf Z}-point (g). }
  \label{z-polar}
\end{figure}

Basing on a simple ionic lattice dynamical model one expects
roughly similar frequencies for the $A_g$ and the \ozz -modes.
However, we find a steep downwards dispersion in the $\Delta
_1$-branch connecting these two modes, see Fig. 6. From the
$A_g$-frequency at 16.4\ THz the frequencies of the branch
decrease towards the value of  14.5\ THz observed for the \ozz
-mode. Two scans across this branch are shown in Fig. 1d). In
contrast the $\Delta _4$ branch exhibits an increasing dispersion
from the $A_u$-mode towards the corresponding {\bf Z}-point mode.
This dispersion cannot be described with the normal ionic model,
the corresponding dispersion is indicated in Fig. 6 by the broken
lines. In order to describe the pronounced softening of the \ozz
-mode without affecting the description of the other modes we had
to introduce two specially adapted force constants, see table I.
The \ozz \ force constant acts on the two O2 sites connected to
the same Ru, and the breathing constant chosen negative acts on
the in-plane oxygens connected to the same Ru. These constants
mimic the observed dispersion rather well but they do not possess
an intrinsic physical meaning. With these additional parameters a
good description of the bond-stretching dispersion is obtained
except the transverse $\Sigma _3$ branch connecting to the
quadrupolar mode at \vq =(0.5,0.5,0), see Fig. 1f) for a raw-data
scan and Fig. 5d) for the polarization pattern, where the model
predicts a softening stronger than the one found experimentally.

\begin{figure}
  \includegraphics[width=0.27\textwidth]{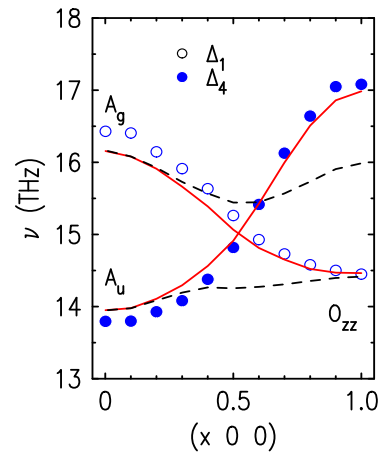}
  \caption{(color online) Dispersion of the z-polarized bond-stretching phonons along the
  [100]-direction; the $A_g$/\ozz \ and $A_u$/{\bf Z}-$A_u$ modes connect through $\Delta _1$
  and $\Delta _4$-branches, respectively. The broken lines give the results of a normal lattice
  dynamics model which is unable to describe the dispersion properly. The solid lines correspond
  to the dispersion calculated with the inclusion of the special force constants.}
  \label{ozz-disp}
\end{figure}

We think that the anomalously soft \ozz -mode frequency has to be
attributed to a strong electron-phonon coupling, as it is further
supported by the temperature dependence. In two independent
experiments we have studied the temperature dependence of these
c-axis polarized $\Delta _1$-modes between 15 and 300\ K and
between 15 and 450\ K. The results are resumed in Fig. 7. The
entire $\Delta _1$-branch softens upon heating, but the softening
is much stronger at the \ozz -mode where it exceeds the
expectation for a typical mode Gr\"uneisen parameter, see Fig.
7b) and c). In addition, at room temperature the \ozz -mode is
significantly broader than the other modes studied in this
branch. Most interestingly the anomalous softening of the \ozz
-mode occurs just above about $\sim$150\ K, which is the crossover
temperature observed in the $c$ axis electric resistivity
\cite{hussey,tyler}. Concomitantly, the width of the \ozz -mode
increases at this temperature range. The measurements of the
optical conductivity in \srruo \ have revealed that the charge
transport along the $c$ direction is coherent only at low
temperatures \cite{katsufuji}.

We  attribute the anomalous dispersion near the \ozz -mode as
well as the temperature effects to a strong coupling between this
phonon mode and the inter-layer charge transport. This electron
lattice coupling causes considerable phonon softening and
broadening in particular at high temperature where the charge
transport is incoherent. Evidence for a much stronger
electron-phonon coupling associated with the \ozz -mode was found
in the isostructural cuprate \lasrcuo \ \cite{5} and in less
extent in \nccoef \ \cite{27}. Ho and Schofield have explained
the temperature dependence of the anisotropic resistivity of a
two-dimensional metal and in particular the existence of a
resistivity maximum along the less coupled direction by an
anisotropic electron-boson coupling \cite{ho}. They obtained an
excellent fit to the $c$-axis resistivity in \srruo \ assuming the
characterisic energy of the coupling boson at $k_B \cdot$515\
K$\sim$11\ THz, which agrees reasonably well with the energy of
the \ozz -mode of 14\ THz, corroborating the conclusion that
strong electron-phonon coupling for the \ozz-mode is associated
with the uncommon c-axis charge transport.

\begin{figure}
\includegraphics[width=0.35\textwidth]{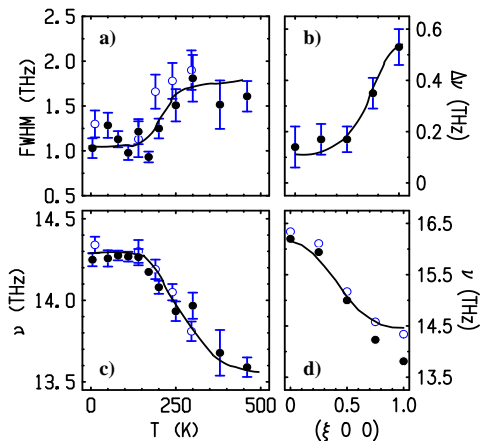}
\caption{(color online) Temperature dependencies of peak widths
and frequencies of the c-polarized bond-stretching $\Delta _1$
modes. a) and c) : temperature dependencies of the peak width (a)
and of the frequency (c)  of the \ozz -mode, respectively. These
results were determined in two independent experiments (open and
filled symbols). b) and d): \vq -dependence of the
temperature-induced frequency shifts (b) and of the phonon
frequencies of this $\Delta_1$ branch (d) at 15\ K (open symbols)
and at room temperature (filled symbols).} \label{agau-temp-fig}
\end{figure}

Bands of different orbital character contribute to the Fermi
surface in \srruo \ \cite{maeno-mackenzie}. Amongst them, the \dxz
\ and the \dyz \ states should be most relevant for the c-axis
charge transport and also most sensitive to the strong coupling
with the \ozz \ phonon mode. The electron-phonon coupling, thus,
should be more important for these \dxz \ and  \dyz \ states
possibly competing with a non-conventional superconducting pairing
mechanism. The anisotropic electron-phonon coupling in \srruo \
might, therefore, enhance the orbital selective character of the
superconductivity in \srruo \ \cite{agterberg}.

\section{Search for a Kohn-anomaly}

Pure \cite{sidis} and Ti-doped \srruo \ \cite{tisrruo} are
text-book examples for pronounced Fermi-surface nesting. The
quasi-one dimensional \dxz \ and \dyz \ bands form flat
Fermi-surface sheets yielding strong nesting at the \vq -position
(0.3,0.3,$q_l$). In pure \srruo \ this nesting implies sharply
peaked magnetic fluctuations \cite{sidis,braden2002,servant}, and
upon minor Ti-substitution a static spin-density wave develops at
this propagation vector \cite{tisrruo}. We have studied the
question whether the pronounced peak in the electronic
susceptibility yields a signature in the lattice dynamics as well
which would correspond to a Kohn anomaly.

The longitudinal breathing modes along the [110] direction are a
first promising candidate to search for a Kohn anomaly, since the
breathing modes are intrinsically coupled with the Ru-site
charges. In the in-plane bond stretching dispersion shown in Fig.
4, however, no such signature can be detected in spite of the high
precision obtained for that branch. Another candidate for a Kohn
anomaly is the longitudinal acoustic branch which belongs to the
same $\Sigma _1$ representation. In order to detect such an
anomaly we have measured the phonon-frequencies in the lowest two
$\Sigma _1$ branches with much higher precision than the data
taken in order to establish the global dispersion in Fig. 2. At
60\ K one data set was taken in the [100]/[010] scattering
geometry using the PG monochromator. We also used the [110]/[001]
scattering geometry in combination with the PG monochromator
yielding comparable results but less precision due to the less
favorable alignment of the resolution ellipsoid. Another data set
was taken in [100]/[010] scattering using the Cu-(111)
monochromator yielding a better energy resolution at 12 and 295\
K. The results of these measurements are shown in Fig. 8. Indeed
one may discern a small dip in the lowest branch just at \vq
=(0.3,0.3,0) which is somehow smeared out at room temperature.
This dip, however, is superposed with the effects arising from
the anti-crossing behavior of the two branches of the same
symmetry which may explain the weak plateau in the acoustic
branch around \vq =(0.25,0.25,0)-(0.35,0.35,0). Anyway the effect
of the nesting has to be considered as being weak. It appears
very unlikely that such an electron phonon coupling is a relevant
effect in \srruo .

\begin{figure}
\includegraphics[width=0.42\textwidth]{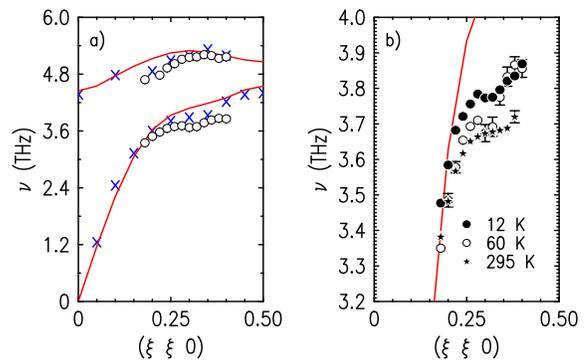}
\caption{(color online) Dispersion of the longitudinal acoustic
and the lowest optical branch of the same symmetry along the
[110]-direction, $\Sigma _1$-modes. In part a) crosses denote the
data taken to establish the overall dispersion shown in Fig. 2.
Open circles denote the results of measurements at 60\ K with the
PG monochromator; filled circles and stars present the data
obtained with the Cu-(111) monochromator acchieving higher
resolution. Lines correspond to the frequencies calculated with
the lattice dynamical model. } \label{Model}
\end{figure}

\section{Conclusions}

The lattice dynamics of \srruo \ in spite of its excellent
metallic properties at low temperatures still shows the
characteristic features of an ionic compound. The dispersion is
best described in an ionic shell model taking account of the
electronic screening. Compared to previously studied perovskites,
in particular the high-temperature superconducting cuprates, the
screening is more efficient in \srruo . There is no direct
relation between the screening and electric transport properties,
as, in contrast to the transport, the screening is isotropic and
essentially temperature independent.

In view of the anomalous bond-stretching dispersion observed in
all electronically doped perovskites studied so far, this aspect
is particularly interesting in \srruo \ as well. The in-plane
bond-stretching dispersion, however, is rather normal in \srruo .
The anomalous effects in the other materials may hence not simply
be attributed to their metallic character. Instead it seems likely
that the high mobility of the charge carriers along the planes in
\srruo \ inhibits the strong coupling between the breathing modes
and charge fluctuations. The absence of the in-plane
bond-stretching anomalies in \srruo \ together with the
observation of kinks in ARPES spectra resembling those of the
cuprates, renders a dominant role of the in-plane bond-stretching
phonons in the cuprate-kink mechanism unlikely.

A highly anomalous bond-stretching dispersion is, however,
observed for the c-polarized bond-stretching modes in \srruo .
The \ozz \ zone-boundary mode exhibits a rather low frequency and
a strongly temperature-dependent line broadening. These effects
indicate that this mode is coupled with the anomalous charge
transport along the c-direction observed in DC resistivity and in
optical conductivity experiments. Whether this sizeable
electron-phonon coupling plays some role in the superconducting
pairing mechanism in \srruo , remains a challenging open issue.

Work at Universit\"at zu K\"oln was supported by the Deutsche
Forschungsgemeinschaft through the Sonderforschungsbereich 608.

\end{document}